# RANDOM-SITE CATION ORDERING AND DIELECTRIC PROPERTIES OF PbMg$_{1/3}$Nb$_{2/3}$O$_3$-PbSc$_{1/2}$Nb$_{1/2}$O$_3$

I. P. RAEVSKI[1], S. A. PROSANDEEV[1], S. M. EMELYANOV[1], F. I. SAVENKO[1],
I. N. ZAKHARCHENKO[1], O. A. BUNINA[1], A. S. BOGATIN[1], S. I. RAEVSKAYA[1],
E. S. GAGARINA[1], E. V. SAHKAR[1], AND L. JASTRABIK[2]

[1]Research Institute of Physics, Rostov State University, 344090 Rostov on Don, Russia;
[2]Institute of Physics AS CR, Na Slovance 2, 182 21 Prague 8, Czech Republic

(1-$x$)PbMg$_{1/3}$Nb$_{2/3}$O$_3$-($x$)PbSc$_{1/2}$Nb$_{1/2}$O$_3$ (PMN-PSN) solid solution crystals have been grown by the flux method in the whole concentration range. X-ray supercell reflections due to B-cation ordering were observed for as-grown crystals from the $0.1 \leq x \leq 0.65$ compositional range. Though the ordered domains are rather large (~50 nm) the relaxor-like dielectric behavior is observed for compositions with $x < 0.6$. The diffusion of the dielectric permittivity maximum in as-grown crystals is the lowest at $x = 0.6$ and increases towards the end members of solid solution. Such behavior is explained within a Bragg-Williams approach by employing the random layer model. At $x \sim 0.6$ the excitation energy determined from the Vogel-Fulcher relation exhibits a jump which we regard to changing the kind of the polar regions from PbMg$_{1/3}$Nb$_{2/3}$O$_3$ to PbSc$_{1/2}$Nb$_{1/2}$O$_3$ related type.

Keywords    cation ordering, dielectric properties, relaxors

## 1. INTRODUCTION

Solid solution single crystals of some ternary PbB'$_n$B"$_m$O$_3$ perovskites displaying colossal dielectric, piezoelectric, electrostrictive, and electrooptic responces are promising materials for various applications [1]. As the crystal growth takes much more time than sintering of ceramics, it inevitably includes annealing of the inner parts of the crystal, which were formed at higher temperatures. Thus, cation ordering, which is achieved in ceramics via a prolonged heat treatment, is very probable even in as grown crystals. It is well-known that the effect of B-cation ordering is extremely different in 1:1 and 1:2 ternary PbB'$_n$B"$_m$O$_3$ perovskites. A typical 1:1 perovskite PbSc$_{1/2}$Nb$_{1/2}$O$_3$ (PSN) in a highly ordered state exhibits a sharp ferroelectric (FE) phase transition while in a disordered state the FE phase transition is diffused and there is a substantial frequency dependence of the dielectric permittivity $\varepsilon'$ maximum temperature $T_m$. In contrast to this, 1:2 perovskites such as $PbMg_{1/3}Ta_{2/3}O_3$ and some $PbMg_{1/3}Nb_{2/3}O_3$ (PMN) based compositions exhibit a classical relaxor behavior both in the ordered and disordered states. The latter fact was described qualitatively within the random layer model earlier [6,7]. According to this model, the crystal structure of (1-$x$)PMN-$x$PSN (below we will call them also PMN-PSN for short) consists of two alternating (111) planes: the first is made of only Nb, while the second is a random mixture of Mg, Nb and Sc, which can be represented by the first bracket in the chemical formula: $Pb[(Mg_{2/3}Nb_{1/3})_{1-x}(Sc)_x]_{1/2}[Nb]_{1/2}O_3$. In the present study we compare dielectric properties of PMN-PSN as-grown solid solution crystals with those of ordered and disordered ceramics [7].

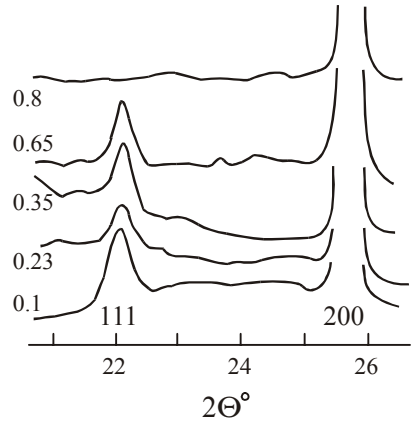

FIGURE 1. Fragments of X-ray diffraction patterns for powdered as-grown (1-$x$)PMN-$x$PSN crystals. Numbers correspond to x values.

## 1. EXPERIMENTAL

PMN-PSN solid solution crystals were grown by the flux method. For all compositions the same flux (PbO-$B_2O_3$), crystallization temperature interval (1150-980 $^0$C) and the cooling rate (10K/h at T>1100 $^0$C and 1K/h in the 1100-980 $^0$C range) were employed. The crystals obtained are transparent, have a yellow color, and an isometric form (the edge dimension is 0.5 mm to 4 mm) with the sides parallel to the (100) planes of the perovskite prototype lattice. The chemical composition of the crystals was determined with the help of the electron probe X-ray microanalizer «Camebax-Micro»; PSN and PMN crystals served as reference samples. X-ray studies of the powdered crystals were carried out at room temperature with the help of DRON-3.0 diffractometer using Co $K_\alpha$ radiation.

Dielectric permittivity $\varepsilon'$ measurements in the $10^{-2}$ to $10^2$ Hz range were carried out in the frequency-domain mode using a computer-controlled experimental setup similar to that described in Ref. [8]. In the $10^2$ to $10^4$ Hz range $\varepsilon'$ was measured with the help of R5083 LCR-meter. Plate-like specimens for dielectric studies were cut parallel to (001) faces and vacuum-evaporated Al was used as the electrode material.

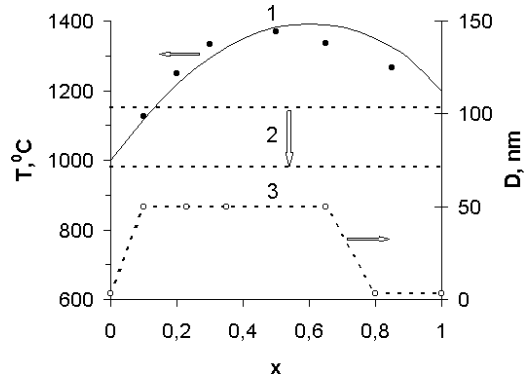

FIGURE 2. The correlation between $T_t$ values (1), crystallization temperature range (2) and ordered domain sizes (3) for (1-$x$)PMN-($x$)PSN crystals. Dashed lines are guides to the eye. Solid curve (1) represents the $T_t(x)$ dependence calculated using expression (6). Filled circles show experimental $T_t$ values [7].

X-ray supercell reflections due to B-cation ordering were observed for as-grown crystals from $0.1 \leq x \leq 0.65$ compositional range (Fig.1). The concentration dependence of ordered domain sizes estimated from the width of the superstructure reflections compared to the fundamental ones is shown in Fig.2. Although the maximal ordered domain sizes are rather large (~50 nm), strongly diffused $\varepsilon(T)$ maximums are observed for all PMN-PSN crystals, especially for those with $x < 0.6$ (Fig. 3). In order to compare the diffusion of $\varepsilon'(T)$ we plot the ratio $\varepsilon'/\varepsilon'_m$ for the studied compositions vs reduced temperature $T-T_m$ where $T_m$ is the temperature of the $\varepsilon(T)$ maximum and $\varepsilon'_m$ is the $\varepsilon'$ maximal value (Fig. 4). One can see that, in PMN-PSN crystals, the diffusion of the dielectric permittivity maximum is the lowest for $x \approx 0.6$ and increases towards the end members of the solid solution. The shape of the dielectric permittivity $\varepsilon$ curve changes substantially at $x > 0.6$, i.e. a sharp step appears on it at temperature $T_s$ somewhat below $T_m$ (Figs. 3 and 4). While $T_m$ increases with frequency, $f$, in accordance with the Vogel-Fulcher relation, the temperature $T_s$ of the step on the $\varepsilon'(T)$ curve is practically independent of $f$. These results are similar to those obtained earlier for annealed PMN-PSN ceramics [7].

Fig. 5 shows the concentration dependence of $T_m$ for PMN-PSN crystals in comparison with the same data for both ordered and disordered ceramics [7]. At $x > 0.6$ the $T_m(x)$ dependences for crystals tend to converge to the $T_m(x)$ curve for disordered ceramics. For PMN-PSN crystals with $0.65 \leq x \leq 0.80$ $T_s$ is close to Vogel-Fulcher temperature, $T_0$, while for PSN crystal $T_s$ is substantially lower than $T_0$.

## 2. DISCUSSION

Previously it was found out that the mean ordering degree $s$ of the as-grown crystals of PSN and some other ternary 1:1 perovskites depend on the relation between the crystallization temperature and the temperature $T_t$ of the compositional order-disorder phase transition, the time of crystallization and the diffusion rate for B-site cations in the temperature interval of crystallization [5, 9-11]. The temperature $T_t$ of compositional order-disorder phase transition for PST-PSN was found to be an approximately linear function of composition [9]. For $(1-x)$PMN-$(x)$PSN system $T_t$ was reported to be a parabolic function of $x$ [7].

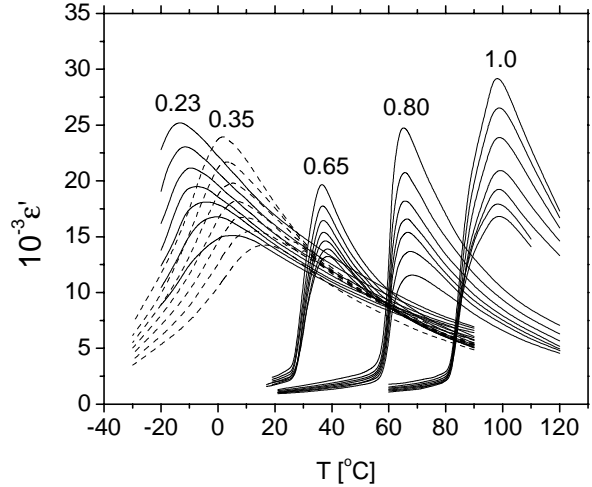

FIGURE 3. $\varepsilon'(T)$ dependencies of $(1-x)$PMN-$x$PSN crystals, measured at different frequencies: $10^{-2}$, $10^{-1}$, $10^0$, $10^1$, $10^2$, $10^3$, $10^4$ Hz.(from top to bottom ). Numbers correspond to $x$ values.

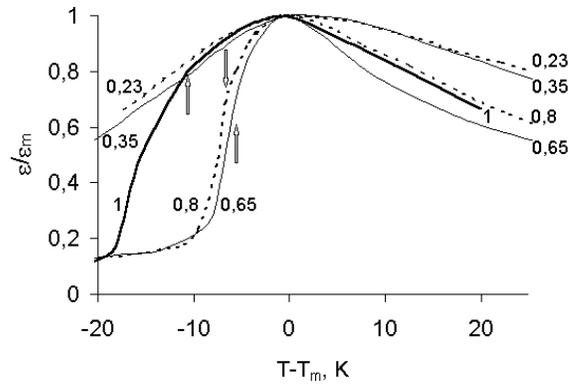

FIGURE 4. Dependence of $\varepsilon/\varepsilon_m$ ($\varepsilon_m$ is the permittivity maximum value) on the reduced temperature $T-T_m$ for $(1-x)$PMN-$(x)$PSN crystals measured at 1 kHz. Arrows show the Vogel-Fulcher temperature..

Below we will derive an analytical expression for the ordering temperature in $(1-x)$PMN-$(x)$PSN and $(1-x)$PST-$(x)$PSN in the framework of the Bragg-Williams approximation [12]. We start with the case of PMN-PSN. The final result in the case of PSN-PST is the same, formally.

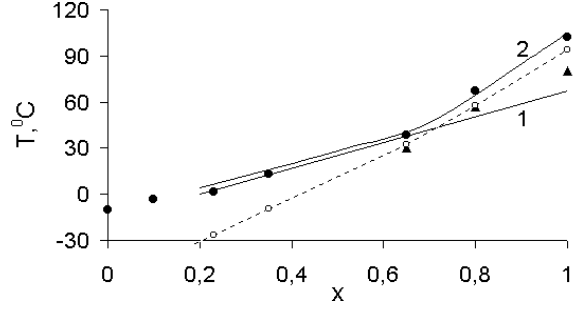

FIGURE 5. The concentration dependence of $T_m$ measured at 1 kHz (filled circles), $T_0$ (open circles) and the temperature $T_s$ of the step on the $\varepsilon'(T)$ curves (filled triangles) for $(1-x)$PMN-$(x)$PSN crystals in comparison with the same data for ordered (1) and disordered (2) ceramics [7].

Consider two sublattices with equal number of sites $N/2$ and three sorts of ions occupying these sites with the probabilities:

$$
\begin{aligned}
p_1^{(1)} &= \tfrac{1}{3}(1-s)(1-x) & p_1^{(2)} &= \tfrac{1}{3}(1+s)(1-x) \\
p_2^{(1)} &= \tfrac{1}{6}(1-s)(1-x) & p_2^{(2)} &= \tfrac{1}{6}(1+s)(1-x) \\
p_3^{(1)} &= \tfrac{1}{2}(1-s)x & p_3^{(2)} &= \tfrac{1}{2}(1+s)x \\
p_4^{(1)} &= \tfrac{1}{2}(1+s) & p_4^{(2)} &= \tfrac{1}{2}(1-s)
\end{aligned}
\quad (1)
$$

Here subscript shows the sort of the ions: 1: $Mg_{2(1-x)/3}$, 2: $Nb_{(1-x)/3}$, 3: $Sc_x$, 4: $Nb_{1/2}$, and the superscript is the sublattice's number. We consider ordering in $Pb[(1-x)(Mg_{2/3}Nb_{1/3})-(x)Sc]_{1/2}[Nb]_{1/2}O_3$ between the ions in the first and second brackets [6,7,13]. The ions in the first bracket are assumed fully disordered (the random layer model). The free energy in the form suggested by Bragg and Williams is

$$F = \sum N_{ij} w_{ij} + k_B NzT \sum p_i (\ln p_i - 1) \quad (2)$$

where $N_{ij} = zN\left(p_i^{(1)} p_j^{(2)} + p_j^{(1)} p_i^{(2)}\right)$, $z$ is the number of nearest neighbors, $s$ is a degree of the order in the considered structure, $w_{ij}$ are pair energies. After substitution (1) to (2) and finding the equilibrium degree of ordering one can obtain

$$T \ln \frac{1-s}{1+s} = s(a + bx - cx^2) \quad (3)$$

where the coefficients $a$, $b$ and $c$ can be expressed in terms of the pair energies $w_{ij}$. The equation obtained allows one to find the dependence of the degree of the order on temperature and concentration. There is an (ordering) temperature, $T_t$, above which the crystal shows disorder. This temperature can be easily found by expanding the logarithm in the series with respect to $s$ and leaving only the first power of $s$:

$$2Ts = s(a + bx - cx^2) \tag{4}$$

Hence the bifurcation point (where a new solution distinct from $s = 0$ appears) is at

$$T_t = \tfrac{1}{2}(a + bx - cx^2) \tag{5}$$

The ordering temperature can be expressed over the corresponding temperatures at $x = 0$ ($T_1$) and $x = 1$ ($T_2$):

$$T_t = T_1 + (T_2 - T_1 + c)x - cx^2 \tag{6}$$

At $c = 0$ the concentration dependence is linear. At $c > 0$ the ordering temperature for intermediate concentrations can be higher than for the boundaries.

Calculations using expression (6) and experimentally estimated values of $T_t \approx 1000\ ^0C$ for PMN [7] and $T_t \approx 1200\ ^0C$ for PSN [2] give a $T_t(x)$ dependence (Fig. 2) which is close enough to experimental one for (1-$x$)PMN-$x$PSN ceramics [7], especially if one takes into account that the boundary obtained in [7] corresponds to some residual order and, hence, the true $T_t$ values are somewhat higher.

During the crystal growth of PSN and PST, the disordered structure is formed first, even when the crystallization temperature is much lower than $T_t$ and the disordered phase is non-equilibrium [5,10,11]. The ordering then starts via a diffusion mechanism. The actual degree of ordering for as-grown crystals depends upon $t/\tau$, where $t$ is the time of crystallization, and $\tau$ the time necessary for achieving equilibrium degree of ordering $s_e$ via diffusion; $\tau$ depends on $T$, while $s_e$ on $T/T_t$.

Our results of X-ray and dielectric studies of the as-grown PMN-PSN crystals are in qualitative agreement with this picture. Superstructure reflections due to B-site ordering are absent in pure PMN having $T_t$ close to the lower boundary of the temperature interval of crystallization, but appear at $x \geq 0.1$, for which $T_t$ exceeds this boundary. Besides the decrease of ordering at large x due to increase of $T/T_t$ ratio (Fig.2), the absence of superstructure reflections for $x \geq 0.8$ is likely to be caused by general lowering of the intensity of these reflections with x due to smaller difference in atomic scattering factors of Sc and Nb as compared to Mg and Nb. (In PSN intensity of superstructure reflections is known to be very low even for high degree of ordering [2]).

Another factor which has an impact on the ordering degree is the rate of ordering at the crystallization temperatures. For (1-$x$)PSN-$x$PST it was established that the rate of ordering for compositions from the PST end of the solid solution system is substantially higher than from the PSN end [9]. The same is true for (1-$x$)PMN-$x$PSN system, as it is evidenced from Fig.5 where high values of $T_m$, corresponding to a low degree of ordering are observed for crystals with $x > 0.65$.

Usually a frequency-independent step on the diffused $\varepsilon(T)$ curve is attributed to a spontaneous transition from the relaxor to the normal ferroelectric state, observed previously in PSN ceramics [14] and crystals [15]. For PMN-PSN crystals with $x = 0.65$ and $x = 0.80$ such an assumption seems justified, because $T_s$ is close to $T_0$, in accordance with data for other complex perovskites [14,15]. One more confirmation is a large drop of excitation energy, determined from the Vogel-Fulcher relation, at $x \approx 0.6$ (Fig. 6). Similar $E_0(x)$ dependence was observed for $(Pb,Ba)Sc_{1/2}Nb_{1/2}O_3$ solid solution crystals at the boundary between relaxor and ferroelectric compositions [15]. In contrast to this, $T_s$ in PSN crystal is substantially lower than $T_0$ (Fig.5), but it is close to $T_m$ for highly-ordered PSN. Besides, there is an inflexion on the $\varepsilon(T)$ curve of this PSN crystal at somewhat higher than $T_s$ temperature which correlates well with $T_0$ (Fig. 4).

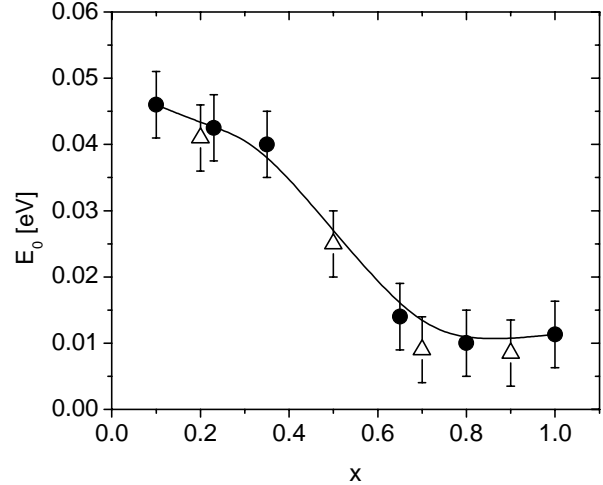

FIGURE 6. Concentration dependence of the Vogel-Fulcher excitation energy $E_0$, for (1-x)PMN-(x)PSN crystals (filled circles) in comparison with the same data for ordered ceramics [7] (open triangles).

It should be noted that, at the crystallization conditions used, the crystals are likely to be ordered inhomogeneously because the internal parts of a crystal have enough time to be ordered while the regions adjacent to the surface, which are formed later, remain disordered. Such inhomogeneity is more probable in PSN-rich compositions because, as was already mentioned above, they have a relatively low rate of ordering. This is consistent with NMR studies of PSN crystals [16]. The step on the diffused $\varepsilon(T)$ curve in PSN (Figs.3 and 4) is likely to be due to the presence of macroscopic highly-ordered regions. Such additional anomalies on $\varepsilon(T)$ dependences have been already observed for PSN crystals grown in a wide temperature interval [10,11].

A relaxor-like dielectric behavior in PMN-PSN ceramics with large ordered domains was earlier described qualitatively within the random-site model of 1:1 B-site cation ordering implying a very short correlation length for ferroelectric coupling due to large random fields in one of the B-sublattices [7,13]. However, some questions remain unsolved. One of them is the nature of the $E_0$ drop at $x \approx 0.6$ (Fig. 6). We suggest that, at this concentration, the kind of the polar region changes from the PMN (or Mg)- related to the PSN (or Sc)- related. In PMN, the closest to Mg cations are strongly shifted to Mg, and oxygen ions are shifted away from Mg; the Pb ions which are at the middle between two nearest Mg's have especially soft dynamics of the Last type due to the frustration of the Pb displacement against the other nearest ions (Nb and O) [17]. The scaling factor for the size $L$ for such a soft quasilocal vibration is just the average distance between two nearest Mg's. In the (1-$x$)PMN-($x$)PSN system the concentration of Mg in the B′-sublattice corresponds to the formula (Mg$_{(2-2x)/3}$Nb$_{(1-x)/3}$Sc$_x$). One can estimate average $L$ as a value proportional to $a[1/(2-2x)]^{1/3}$ in the case of 3D disorder where $a$ is the quasicubic lattice parameter. In the same manner, one can introduce a polar region between two nearest Sc ions: the length $L$ in this case is proportional to $x^{-1/3}$, at a 3D order. This model has two scaling parameters. One can select the first one by taking $L \sim 3$ nm in PMN [13], and by choosing the second scaling parameter so that these two curves intersect at $x = 0.6$. This intersection point, in our opinion, corresponds to the crossover between the polar region's sizes related to PMN and PSN respectively.

Another way to explain the peculiarity of the point $x \approx 0.6$ is to compare the strength of the Mg and Sc ions using their relative charges with respect to the Nb charge (3 for Mg and 2 for Sc). Equating the number of these ions multiplied by the relative charges one can obtain $x = 0.7$, at 3D disorder. This value is close to the experimentally observed value $x \approx 0.6$. At this $x$, the total strength of the Mg and Sc ions is equal: Mg has larger relative charge but smaller population, while Sc has smaller charge but larger population.

In the ordered phase, the random fields are small that results in the appearance of the keen dielectric anomaly observed experimentally at intermediate concentrations. The relatively small magnitude of $\varepsilon'(T)$ at these concentrations can be explained if one assumes that presumably Sc-related polar regions contribute to $\varepsilon'(T)$ at $x > 0.6$, while, at $x < 0.6$, presumably the PMN related polar regions do (see Fig. 3).

We assume that the drop of $E_0$ (Fig. 6) is connected with the fact that the Mg-related polar regions have larger $E_0$ in comparison with the Sc-related polar regions, due to the larger relative charge of Mg. This assumption is consistent with the fact that the temperature peak of dielectric permittivity in PMN-PSN becomes thinner with $x$. Indeed, according to the Debye theory the width of dielectric permittivity temperature peak is proportional to $E_0$ because the scale for the temperature dependence in the Debye exponent is just $E_0$.

The weak temperature dependence of $E_0$ below $x \approx 0.6$ can be explained by increasing Mg-related polar regions (some influence of the lattice parameter increase is also possible) that, according to Chamberlin [18], results in a *decrease* of the activation energy with $x$, because, in this case, internal degrees of freedom are relaxed (in the limit of infinite domains the relaxation is absent at all). The same consequence can be obtained if one takes into account the fact that ordered regions become larger at intermediate $x$ because of the increase of the ordering temperature at these $x$, as we discussed above. The sharp drop of $E_0$ down at $x \approx 0.6$ appears because only smallest domains contribute to dielectric permittivity, the larger domains are broken by smallest ones and do not exist as a whole. We believe that, below $x \approx 0.6$, the smallest are Mg-related polar regions, but, above this concentration, Sc-related polar regions are the smallest.

This assumption can be justified by considering the connection between dielectric susceptibility and polarization correlator in a finite volume $V$ corresponding to the polar region:

$$\chi \sim V^{-2} \int dV dV' <P(r)P(r')> \sim T_c \min(r^2, r_c^2)/\kappa V \qquad (7)$$

where $r_c$ is the correlation radius, $r = (3V/4\pi)^{1/3}$, $\kappa$ is the coefficient in the Landau expansion in front of the gradient term $(\nabla P)^2/2$, and $T_c$ is an extrapolated local Curie temperature; the correlator is described by the Ornstein-Zernike correlation function. It follows from (7) that susceptibility, above $T_c$, increases with $r_c$ quadratically and decreases with temperature as $1/a(T-T_c)$, where $a$ is constant. When $r_c$ exceeds the polar region size $r$ then the considered contribution saturates (the saturation temperature is $T_s = 1/r^2 a + T_c$) and does not further depend on the correlation radius and, hence, on temperature. In reality, of course, this crossover is smoothed due to boundary conditions.

The correlation of polarization between nearest polar regions can be valuable only in ordered samples in which these regions resonate or just at the phase transition temperature. In the case of strong disorder, random fields work against the interaction between nearest polar regions [19]. Hence, in general, one should consider the following different cases when using (7): i) **r** and **r**' belong to the smallest polar regions (Mg related at $x < 0.6$, and Sc related, at $x > 0.6$); ii) **r** and **r**' belong to different smallest polar regions; iii) **r** and **r**' belong to polar region and dielectric interface respectively; iv) **r** and **r**' belong to the dielectric interface.

In the disordered samples, well above the phase transition temperature, only the first contribution is significant We have shown above that this contribution can be described by a Curie-Weiss law at high temperatures, at which the correlation radius is smaller than the radius of a polar region, but this contribution saturates at lower temperatures because of the finite size of polar regions; the local polarization can appear in this case due to local random fields. Just at the phase transition temperature, the interaction of the polar regions should be taken into account in order to describe the cooperative effect [16]: in the case if the dispersion of this interaction is stronger than the average interaction then a glass type phase transition occurs; otherwise, a ferroelectric phase transition takes place.

According to the present model, there are two levels of organization in inhomogeneous ferroelectrics. The first is connected with the appearance of local slowly oscillating polarization in the polar regions. It happens when the correlation radius of ferroelectric fluctuations reaches the heterogeneity size. The next level is the organization of the polar regions themselves. It was described earlier within a random-bond random-field model [16]. Respectively, in relaxors, there are two different order parameters with different critical temperatures. The first order parameter corresponds to the local polarization inside the polar regions. In our opinion, the local polarization in neighbored polar regions appears cooperatively in order to decrease the depolarization field [20]. The other order parameter can be of the glass-type or ferroelectric as it was described above [16].

When adding $PbTiO_3$ to PMN, the polar region size increases and, finally, the phase transition becomes normal ferroelectric. In the case when PMN is mixed with PSN, one simply substitutes the Mg-related polar regions by Sc-related ones. In this case, one should take into account the difference in the Vogel-Fulcher activation energy in the Mg- and Sc-related polar regions, as well as the change of the phase transition temperature due to the change of the lattice parameter (the lattice parameter increases with $x$ in PMN-PSN). Thus, experiments on ternary perovskites performed help to understand the nature of the relaxor state and its development.

## 3. SUMMARY

Basing on the Bragg-Williams approximation in the alloy theory we explicitly obtained an analytical expression for the temperature $T_t$ of the compositional order-disorder phase transition for solid solutions giving all the observed types of $T_t(x)$ dependences. According to model [10], a parabolic shape of the $T_t(x)$ dependence for $(1-x)$PMN-$x$PSN system results in a higher ordering degree of the ceramics at intermediate concentrations in comparison with the boundary ones. According to our experiments, the same regularities of cation ordering during the crystal growth are valid for solid solution crystals of both 1:1 and 1:2 ternary perovskites.

In order to understand the peculiarities of PMN-PSN dielectric behavior deeper we have developed a model according to which, at $x \approx 0.6$, the kind of the polar region changes from the PMN (or Mg)- related to the PSN (or Sc)- related. We define the polar region as the lead ion being at the middle between Mg (Sc) ions and vibrating against its nearest Nb and O ions. The change of the kind of the polar region at $x \approx 0.6$ is due to the intersection of the sizes of the Sc-related and Mg-related polar regions at this point.

## 4. ACKNOWLEDGMENT

This study was partially supported by Russian Foundation for Basic Research (grants # 01-03-33119, 01-02-16029 and 02-02-17781) and LN00A015 of the MSMT CR.